\documentclass[aps,prd,showpacs,showkeys,twocolumn,superscriptaddress,groupedaddress]{revtex4}  

\usepackage{graphicx}  
\usepackage{dcolumn}   
\usepackage{bm}        
\usepackage{amssymb}   
\usepackage{latexsym,bm}
\usepackage[tbtags]{amsmath}

\begin{document}

\widetext


\title{Kepler's third law of n-body periodic orbits in a Newtonian gravitation field}

\author{Bohua Sun}

\affiliation{Cape Peninsula University of Technology, Cape Town, South Africa\\
email:sunb@cput.ac.za}%



\begin{abstract}
\small
This study considers the periodic orbital period of an n-body system from the perspective of dimension analysis. According to characteristics of the n-body system with point masses $(m_1,m_2,...,m_n)$, the gravitational field parameter, $\alpha \sim Gm_im_j$, the n-body system reduction mass $M_n$, and the area, $A_n$, of the periodic orbit are selected as the basic parameters, while the period, $T_n$, and the system energy, $|E_n|$, are expressed as the three basic parameters. Using the Buckingham $\pi$ theorem, We obtained an epic result, by working with a reduced gravitation parameter $\alpha_n$, then predicting a dimensionless relation $T_n|E_n|^{3/2}=\text{const} \times \alpha_n \sqrt{\mu_n}$ ($\mu_n$ is reduced mass). The const$=\frac{\pi}{\sqrt{2}}$ is derived by matching with the 2-body Kepler's third law, and then a surprisingly simple relation for Kepler's third law of an n-body system is derived by invoking a symmetry constraint inspired from Newton's gravitational law: $T_n|E_n|^{3/2}=\frac{\pi}{\sqrt{2}} G\left(\frac{\sum_{i=1}^n\sum_{j=i+1}^n(m_im_j)^3}{\sum_{k=1}^n m_k}\right)^{1/2}$. This formulae is, of course, consistent with the Kepler's third law of 2-body system, but yields a non-trivial prediction of the Kepler's third law of 3-body: $T_3|E_3|^{3/2}= \frac{\pi}{\sqrt{2}} G \left[\frac{(m_1m_2)^3+(m_1m_3)^3+(m_2m_3)^3}{m_1+m_2+m_3}\right]^{1/2}$. A numerical validation and comparison study was conducted. This study provides a shortcut in search of the periodic solutions of three-body and n-body problems and has valuable application prospects in space exploration.
\end{abstract}



\maketitle

\begin{figure}[h!]
\centerline{\includegraphics[scale=0.4]{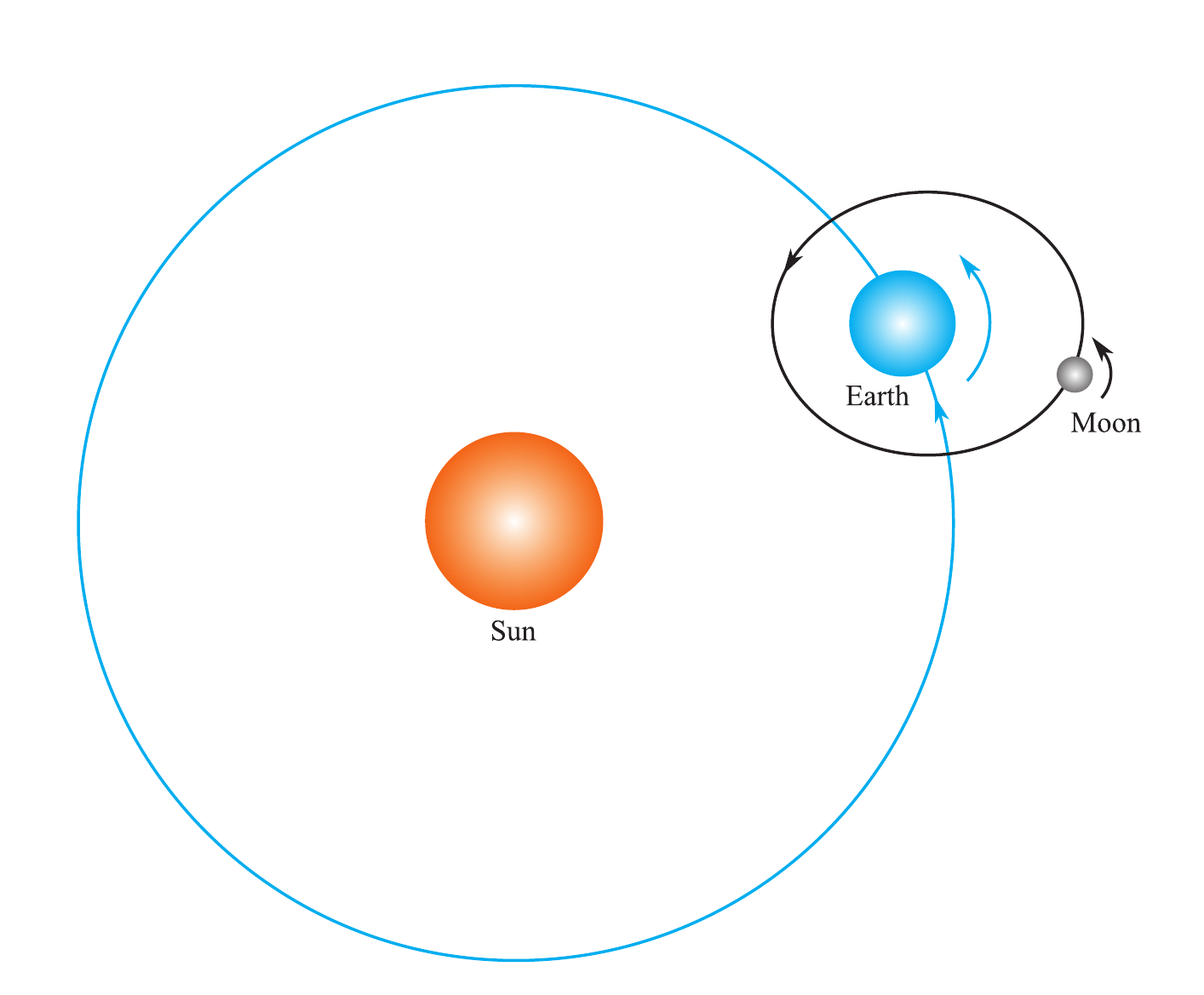}}
\caption{\label{fig01c-1} The Sun-Earth-Moon system}
\end{figure}

\begin{figure}[h!]
\centerline{\includegraphics[scale=0.6]{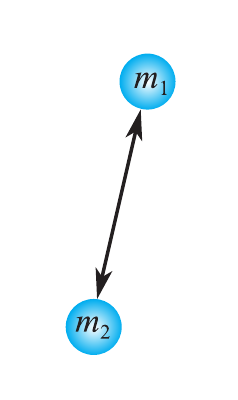}}
\caption{\label{fig01c-2} 2-body system}
\end{figure}

One of the central and most vivid problems of celestial mechanics in the 18th and 19th centuries was the motion description of the Sun-Earth-Moon system under the Newtonian gravitation field (Fig.\ref{fig01c-1}). 
Notable work was done by Euler (1760), Lagrange (1776), Laplace (1799), Hamilton (1834), Liouville (1836), Jacobi (1843), and Poincar\'e (1889) \cite{poi} and Xia (1992) \cite{xia}. The study of the motion between the two bodies was solved by Kepler (1609) and Newton (1687) early in the 17th century. For the elliptic periodic orbit of 2-body system, Kepler's third law of the two-body system \cite{lan} is given by $T|E|^{3/2}=\frac{\pi}{\sqrt{2}} Gm_1m_2\sqrt{\frac{m_1m_2}{m_1+m_2}}$, where the gravitation constant, $G=6.673\times 10^{-11} m^3kg^{-1}s^{-2}$, the orbit period, $T$, the total energy of the 2-body system, $|E|$, and point masses $m_1$ and $m_2$ (Fig.\ref{fig01c-2}).

\begin{figure}[h!]
\centerline{\includegraphics[scale=0.6]{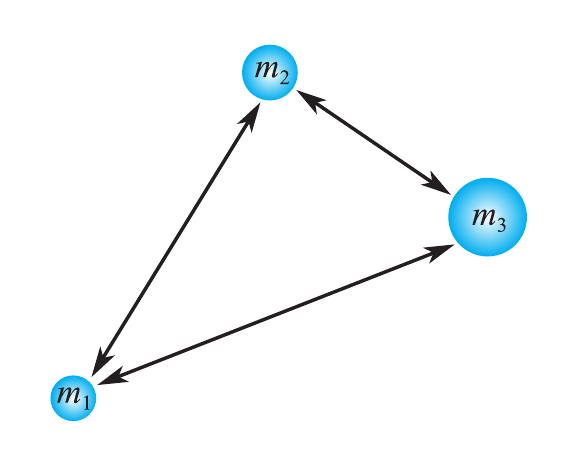}}
\caption{\label{fig01c-3} 3-body system}
\end{figure}

However, the 3-body system (Fig.\ref{fig01c-3}) cannot be solved analytically because unlike the 2-body problem, the 18 variables that describe the system cannot be reduced to a single variable. Simplification of the two-body problem was allowed by invariance and conserved quantities as 'first integrals'. It was proven impossible to reduce the 18 variables of the 3-body problem in order to produce an analytic solution.

Notwithstanding that the analytic solution cannot be found, it is possible to find a numerical solution for the 3-body problem, in which the study of the periodic 3-body orbit has received particular attention in recent years \cite{moo,che,suv-1,dmi-1,dmi-2,suv-2,liao-1,liao-2}. The figure-eight orbit was discovered numerically in 1993, using the principle of the least action by C. Moore \cite{moo}. Chenciner and Montgomery \cite{che} rediscovered the solution by using the shape space least action principle in 2000. In 2013 \v{S}uvakov and V. Dmitra\v{s}inovi\'c \cite{suv-1} made a breakthrough and produced epic results, they found 13 new distinct collisionless periodic orbits of the Newtonian planar 3-body system with an equal mass and zero angular momentum. In 2017 Li and Liao \cite{liao-1}, Li, Jing and Liao \cite{liao-2} reported their breakthrough new finding: 695 periodic orbits of planar 3-body system with an equal mass and zero angular momentum, as well as 1223 periodic orbits of the planar 3-body problem with an unequal mass and zero angular momentum, respectively.

Corresponding to the new finding of more and more periodic obits, a fundamental conjecture was proposed by numerical experiments \cite{suv-2,liao-1,liao-2}, which states that the 3-body system $(m_1,\,m_2\,m_3)$  may obey a law, which is similar to the law of harmonies, named Kepler's third law. Analogous to the Kepler's third law of the two-body problem, \v{S}uvakov and V. Dmitra\v{s}inovi\'c \cite{suv-2} proposed a generalized 3-body Kepler's third law: $T|E|^{3/2}=\text{constant}$, where $|E|$ denotes the total kinetic and potential energy of the 3-body system, where $T$ is the period of periodic orbit. However, they pointed out that "the constant on the right-hand-side of this equation is not ¡°universal¡± in the 3-body case, as it is in the two-body case, and it may depend on both the family of the 3-body orbit and its angular momentum" \cite{suv-2}. Li and Liao \cite{liao-1} enhanced the aforementioned relation to $\bar{T}|E|^{3/2}=\bar{T}^*$, and numerically proved that $\bar{T}^*$ is approximately a universal constant, namely $\bar{T}^*=2.433\pm 0.075$, for the 3-body system with an equal mass and zero angular momentum. This remarkable scale-invariance period $\bar{T}^*=\bar{T}|E|^{3/2}$ was again proved by Li, Jing and Liao \cite{liao-2,com-1} for the 3-body system with unequal mass and zero angular momentum, where $\bar{T}^*=3.074 m_3- 0.617$ in the case of $m_1=m_2=1$ and $m_3$ varied.

Although the aforementioned relations \cite{suv-2,liao-1,liao-2} were supported by the numerical experiments in Ref.\cite{suv-2,liao-1,liao-2}. The questions still remain whether $T|E|^{3/2}=\text{constant}$ is universal, and if not, what form it will takes and how to formulate it without a further numerical simulation. Would we be able to find similar relations or universal scaling laws, for an n-body system, and, if so, how?

The 2-body system incorporates Kepler's three laws, name the law of ellipses, the law of equal areas and the law of harmonies. Clearly, for the 3-body system, since the periodic orbit is no longer elliptical, so that there is no corresponding law of ellipses, and because the periodic orbital topology is more complex, the law of equal areas might also not established. From a large number of numerical simulations of the 3-body system \cite{suv-2,liao-1,liao-2}, the time of each object walking along its orbit is the same, that is, for a given mass of the 3-body system, the periodicity of the periodic orbit in the gravitational field might satisfies Kepler's third law, namely the law of harmonies.

The question now is whether or not the conclusions drawn from those limited numerical experiments are prevalent in the 3-body system. If so, can it possibly be extended to an n-body system? Clearly, with an increase in the number of point mass, the system will have more and more degrees of freedom; while accordingly, the dynamic process becomes hugely complex. If you continue to use numerical simulation to study an n-body problem, the calculation will certainly become more and more challenging. Hence, we have no choice but to find an alternative approach.

This study has attempted to attack an n-body system (including 3-body) by using dimensional analysis \cite{bri,sed,sun-3,hec}. The most powerful use of dimensional analysis is to predict the outcome of an numerical experiment, depending on the variables, whilst providing theoretical insight. Dimensional analysis may come across as simply trying to fit pieces of a puzzle together by trial and error. However, identifying the quantities that are relevant for a given problem is a demanding task, which requires deep physical insight \cite{hec}. This may be done as follows: make a list of all quantities on which the answer must depend, then write down the dimensions of these quantities, and finally demand that these quantities should be combined into a functional form that provides the right dimension. This scheme was cast into a formal framework by Buckingham in 1921 and is often referred to as the Buckingham $\pi$-theorem \cite{buc}.

An important class of central gravitation fields is formed by those in which the potential energy is inversely proportional to the radius $r$, and the force accordingly inversely proportional to $r^2$. They include the fields of Newtonian gravitational and of Coulomb electrostatic interaction; the latter may be either attraction or repulsive. For the attractive gravitation field, the potential energy is $U(r)=-\alpha/r$, where $\alpha=Gm_1m_2$ is a positive constant with dimension $[L]^3[M][T]$ \cite{lan}.

\begin{figure}[h!]
\centerline{\includegraphics[scale=0.6]{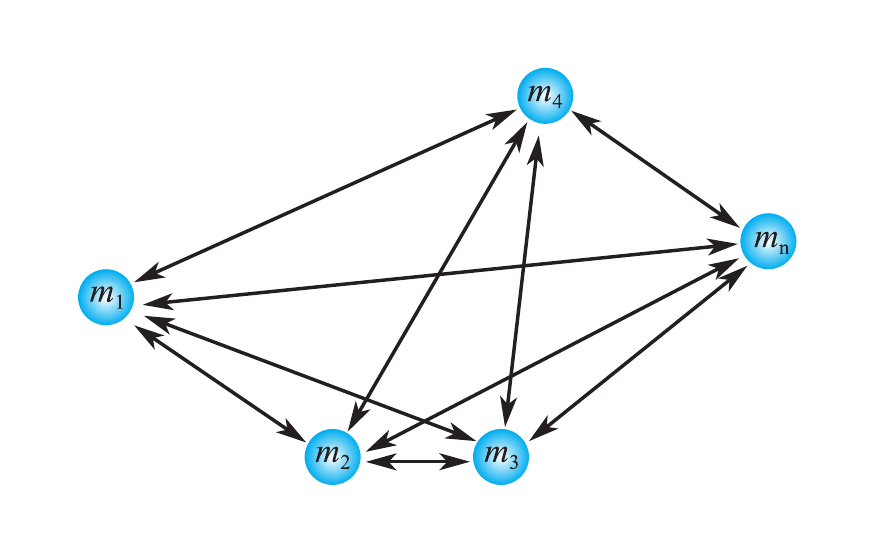}}
\caption{\label{fig01c-n} n-body system}
\end{figure}

Let us have an n-body system with $n$ point masses, denoted by $m_k \,(k=1,\ldots, n)$ (Fig.\ref{fig01c-n}), where each mass does periodic orbital motion in the Newtonian gravitational field, whose gravitational constant is $G$. Assuming that each point mass has no angular rotation and no collision with each other. The question becomes how to extract the basic parameters of the problem from these limited information. The success or failure of using dimensional analysis depends on how to select the basic parameters of the problem. If one choose the wrong parameters, it will lead to absurd results. Hence, the question is, what are the basic parameters for an n-body system?

In physics, these basic parameters should include the gravitational field, the characteristic area or length scale of the periodic orbit, as well as the characteristic mass of the system. It is clear that we can use the gravitational constant $G$ to describe the Newtonian gravitational field; the characteristic mass can be reduced mass $\mu_n$. Each cycle of the track is different, and not like the elliptical orbit of the semi-long axis, which is the characteristic scale. The periodic orbit has a common topological feature, which is that all periodic orbit are closed, hence the area $ A_n$ of the closed orbit could be chosen as the characteristic area scale of the orbit. Its square root is the length scale. Now the periodic orbit problem of an n-body system is to find the orbital period, $T_n$, and total energy, $|E_n|$, which is the summation of kinetic and potential energy. Here it takes its abstract value since it is negative for periodic orbit. From Newtonian gravitation theory, the attraction forces between bodies are linear proportional to the product of $Gm_im_j$. This means that the gravitation constant $G$ can be absorbed into a new parameter $\alpha_n$, whose dimension is $[L]^3[M][T]^{-2}$. In the following, we will use $\alpha$ instead of $G$ as basic parameter. The dimensions of those parameters are listed in Table 1.
\begin{table}[h]
\caption{Parameters and Dimensions}\label{tab1}
\footnotesize
\centerline{
\begin{tabular}{c|c|c|c|c}
\hline
$\alpha_n$ & $\mu_n$ & $A_n$ & $T_n$ &$|E_n|$\\
\hline
$[L]^3[M][T]^{-2}$ & $[M]$ & $[L]^2$ & $[T]$ & $[M][L]^2 [T]^{-2}$\\
\hline
\end{tabular}}
\end{table}

According to the dimensional analysis \cite{bri,sed,sun-3}, the total energy $|E_n|$ can be expressed as
\begin{equation}\label{q1}
  |E_n|=f(\alpha_n,\mu_n,A_n),
\end{equation}
where $f$ stands for a function. This relation has four parameters with three basic dimensions, namely time $[T]$, mass $[M]$ and length $[L]$. From Buckingham $\pi$-theorem \cite{bri,sed,sun-3}, it can produce only one dimensionless parameter, $\pi=|E_n| \alpha_n^a\mu_n^bA_n^c$, the homogenous dimension theorem gives us $a=-1,\, b=0,\, c=1/2$, hence $\pi=|E_n|A_n^{1/2}/\alpha_n$, which must be a constant since it is one only. Therefore, we have
\begin{equation}\label{q2}
  |E_n|A_n^{1/2}(\alpha_n)^{-1}=\text{const.}
\end{equation}
In a similar way, the orbital period $T_n$ can be expressed as $T_n=F(\alpha_n,M_n,A)$, where $F$ stands for a function. This relationship has four variables with three basic dimensions, namely time $[T]$, mass $[M]$ and length $[L]$. It can produce only one dimensionless parameter, $\pi=T_nA^{-3/4}\alpha_n^{1/2}\mu_n^{-1/2}$, which must also be a constant. Therefore n-body Kepler's third law is given by
\begin{equation}\label{q3}
  T_nA_n^{-3/4}\alpha_n^{1/2}\mu_n^{-1/2}=\text{const.}
\end{equation}
Combining the Eq.(\ref{q2}) and(\ref{q3}) and removing the area $A_n$, we can obtain a popular format of the Kepler's third law as follows
\begin{equation}\label{q4}
  T_n|E_n|^{3/2}=\text{const.} \times \alpha_n \sqrt{\mu_n}.
\end{equation}
It is noted that the period depends on the energy of the mass. The higher energy the system has, the shorter period it has. For each energy level, there is a corresponding period, therefore infinite periodic orbits exist \cite{liao-1,liao-2}.

Although Eq.(\ref{q4}) has been formulated, we still cannot get much useful information if the constant, $\alpha_n$ and $\mu_n$ cannot be defined. In other words, the success and failure of Eq.(\ref{q4}) totally depends on determination of these three parameters. The current situation is even more difficult since the only analytic information is the 2-body Kepler's third law. Let us embark on our journey from Kepler's third law.

Since dimensional result Eq.(\ref{q4}) is a general result and should also be true for 2-body system, we can determine the constant by this understanding. In Eq.(\ref{q4}), if we can set the reduced mass $\mu_2=\frac{m_1m_2}{m_1+m_2}$ and parameter $\alpha_2=Gm_1m_2$, by comparing with Kepler's third law $T|E|^{3/2}= \frac{\pi}{\sqrt{2}} G \left[\frac{(m_1m_2)^3}{m_1+m_2}\right]^{1/2} $, then we can propose that const.=$\frac{\pi}{\sqrt{2}}$.

If we carry on this process and will face a big challenge, that is how to extend the 2-body Kepler's third law to an n-body system? From Newtonian gravitation theory, for 2-body system $(m_1,m_2)$, the attraction forces in between are proportional to the linear combination of $m_1m_2$; for 3-body system $(m_1,m_2,m_3)$, the attraction forces between bodies are proportional to the linear combination of $m_1m_2$, $m_1m_3$ and $m_2m_3$; and for 4-body system $(m_1,m_2,m_3,m_4)$, the attraction forces between bodies are proportional to the linear combination of $m_1m_2$, $m_1m_3$, $m_1m_4$, $m_2m_3$, $m_2m_4$, and $m_3m_4$; and for 5-body system $(m_1,m_2,m_3,m_4,m_5)$, the attraction forces between bodies are proportional to the linear combination of $m_1m_2$, $m_1m_3$, $m_1m_4$, $m_1m_5$, $m_2m_3$, $m_2m_4$, $m_2m_5$, $m_3m_4$, $m_3m_5$, and $m_4m_5$; and so on, for an n-body, the attraction forces between bodies are proportional to the linear combination of $m_im_j$, $i=1...n-1, j=i+1$. From mathematics of combination, the number of combination of mass product is $(\begin{array}{c}
                                                                   n \\
                                                                   2
                                                                 \end{array})=\frac{n!}{2(n-2)!}$.

We know that Kepler's third law $T_2|E_2|^{3/2}= \frac{\pi}{\sqrt{2}} G \left[\frac{(m_1m_2)^3}{m_1+m_2}\right]^{1/2} $, which has only one mass product $m_1m_2$ of 2-body system. However, for 3-body system we have three mass product combinations, namely $m_1m_2,\,m_1m_3,\,m_2m_3$, an analogy to Kepler's third law, let us propose 3-body Kepler's third law as follows
\begin{equation}\label{q}
   T_3|E_3|^{3/2}= \frac{\pi}{\sqrt{2}} G \left[\frac{(m_1m_2)^3+(m_1m_3)^3+(m_2m_3)^3}{m_1+m_2+m_3}\right]^{1/2}.
\end{equation}
Clearly, when $m_3=0$, Eq.(\ref{q}) is reduced to 2-body Kepler's law. In physics, any n-body Kepler's law should be able to give 2-body Kepler's law, which means that the n-body Kepler's third law must be compatible with 2-body Kepler's law. In this regards, our formulation Eq.(\ref{q}) is clearly compatible with Kepler's third law.

In the light of Kepler's third law, applying symmetry of mass product in Newtonian gravitation field, we would like to propose following conjecture: an n-body Kepler's third law could be expressed as follows
\begin{equation}\label{q5}
  T_n|E_n|^{3/2}=\frac{\pi}{\sqrt{2}} G\left(\frac{S_n}{M_n}\right)^{1/2}.
\end{equation}
where $S_n=\sum_{i=1}^n\sum_{j=i+1}^n(m_im_j)^3$ and total mass $M_n=\sum_{k=1}^n m_k$. Eq.(\ref{q5}) has answered the conjecture proposed by \cite{suv-1}, namely, $T|E|^{3/2}=\text{constant}$, the constant on the right-hand-side of this equation for a specific mass system is only a constant rather than ¡°universal¡±. For instance, $S_4=(m_1m_2)^3+(m_1m_3)^3+(m_1m_4)^3+(m_2m_3)^3+(m_2m_4)^3+(m_3m_4)^3$, and $S_5=(m_1m_2)^3+(m_1m_3)^3+(m_1m_4)^3+(m_1m_5)^3+(m_2m_3)^3+(m_2m_4)^3+(m_2m_5)^3+(m_3m_4)^3+(m_3m_5)^3+(m_4m_5)^3$.

To compare with numerical simulation results carried out by \cite{liao-1,liao-2}, numerical validation will be conducted for following cases.

Case 1: For 2-body system with  $G=1$ and $m_1=m_2=1$, Kepler's third law gives $T_2|E_2|^{3/2}=\frac{\pi}{2}=1.5707963$.

\begin{figure}[h]
\centerline{\includegraphics[scale=0.3]{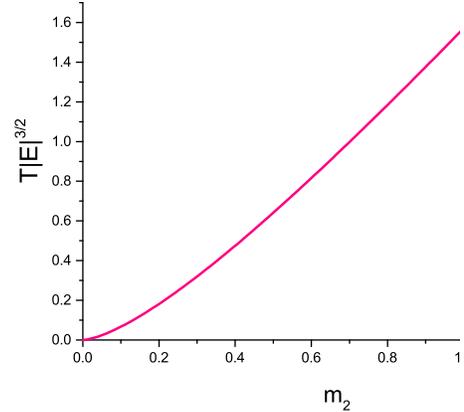}}
\centerline{\includegraphics[scale=0.3]{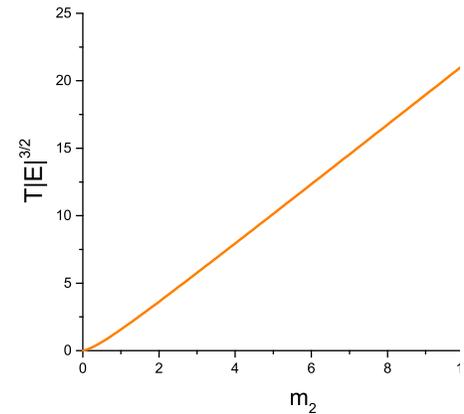}}
\caption{\label{fig2body} Kepler's law for $G=1$ and $m_1=1$ and $m_2$ varied, left is for $m_2 \in [0,1]$, and right is for $m_2 \in [0,10]$.}
\end{figure}

Case 2: For 2-body system with $G=1$ and $m_1=1$ and $m_2$ varied. Kepler's third law gives $T_2|E_2|^{3/2}=\frac{\pi}{\sqrt{2}}(\frac{m_2^3}{1+m_2})^{1/2}$. This law is plotted in Fig \ref{fig2body} for different range of mass $m_2$. The law has obvious nonlinearity at $m_2<1$ and will be more linear as $m_2$ increasing.

Case 3: For 3-body system with $G=1$ and $m_1=m_2=m_3=1$, Li and Liao \cite{liao-1} obtained $ T|E|^{3/2}=2.433\pm 0.075$ through numerical curve-fitting. In this case,  Eq.(\ref{q}) produces $ T_3|E_3|^{3/2}=\frac{\pi}{\sqrt{2}}=2.22144$. Our prediction is close to 2.358 with error $5.8\%$.

\begin{figure}[h]
\centerline{\includegraphics[scale=0.3]{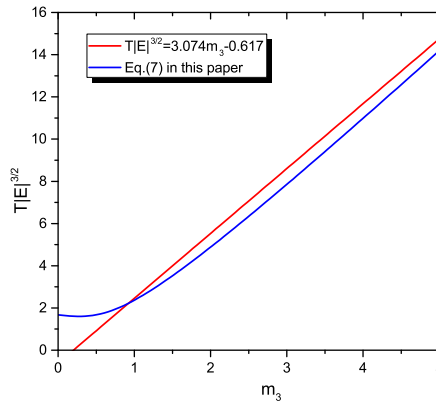}}
\centerline{\includegraphics[scale=0.3]{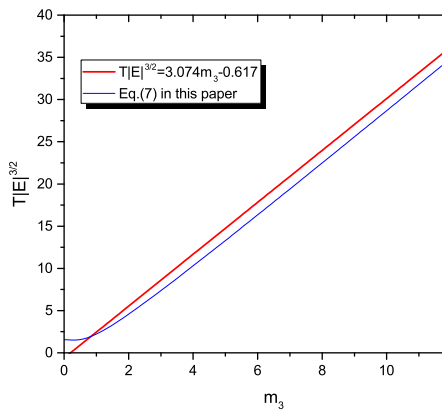}}
\caption{\label{fig0} Comparing for different $m_3$, left is for $m_3 \in [0,5]$, and right is $m_3 \in [0,12]$.}
\end{figure}

\begin{figure}[ht]
\centerline{\includegraphics[scale=0.3]{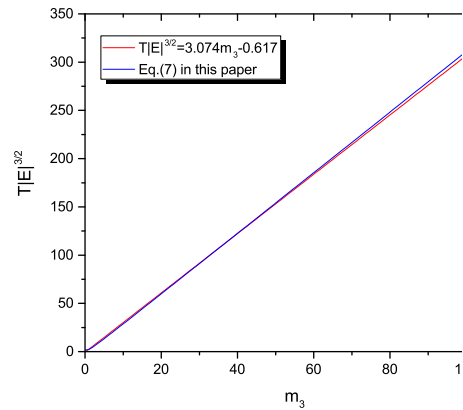}}
\centerline{\includegraphics[scale=0.3]{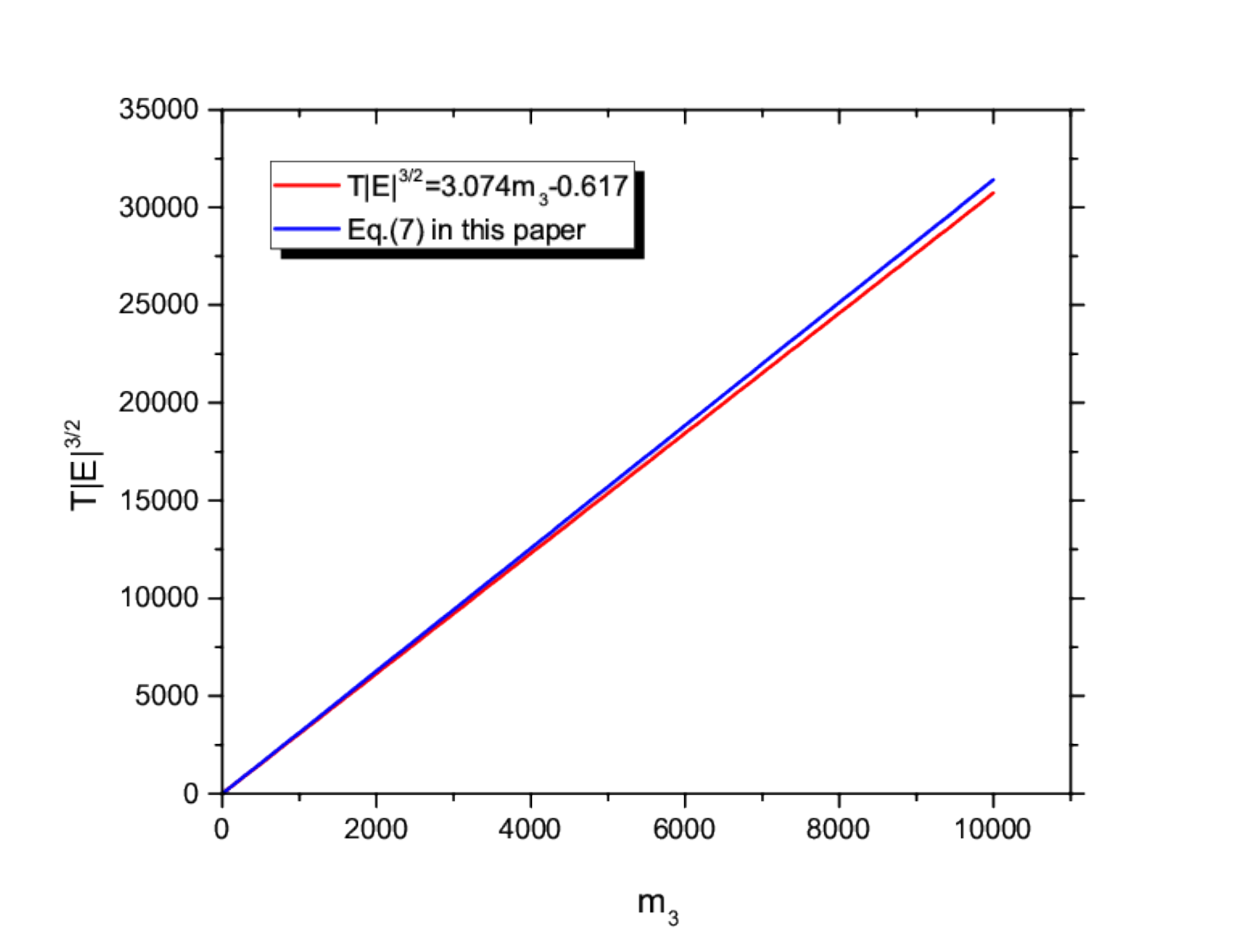}}
\caption{\label{fig1} Comparing for different $m_3$, left is for $m_3 \in [0,100]$, and right is for $m_3 \in [1,10000]$.}
\end{figure}

Case 4: For 3-body system with $G=1$ and $m_1=m_2=1$ and $m_3$ varied., Li, Jing and Liao's \cite{liao-2} proposed linear law $T|E|^{3/2}=3.074m_3-0.617$. In this case, Eq.(\ref{q}) gives
\begin{equation}\label{q7}
   T_3|E_3|^{3/2}=\frac{\pi}{\sqrt{2}} \left[\frac{1+2(m_3)^3}{2+m_3}\right]^{1/2}.
\end{equation}
This relation indicates the Kepler's third law is nonlinear function of $m_3$. The numerical comparing are illustrated in Fig.\ref{fig0} and Fig.\ref{fig1}.

The difference for $m_3\in [0,1]$ shown in Fig.\ref{fig0}, might be interpreted as follows: If set $m_3=0$, Eq.(\ref{q7}) gives 2-body Kepler's third law $ T_2|E_2|^{3/2}=\frac{\pi}{2}$; however, the linear law  $T|E|^{3/2}=3.074m_3-0.617$ obtained by Li, Jing and Liao \cite{liao-2} gives $T|E|^{3/2}=-0.617$, which might have no physical meaning in domain [0,1], since $T|E|^{3/2}$ should be positive. Of course, the linear law  $T|E|^{3/2}=3.074m_3-0.617$ is valid in domain $[1,\infty]$. Generally speaking, the figures show that our formulation has good linearity and keep same trends as the linear law in \cite{liao-2}.

\begin{figure}[h]
\centerline{\includegraphics[scale=0.36]{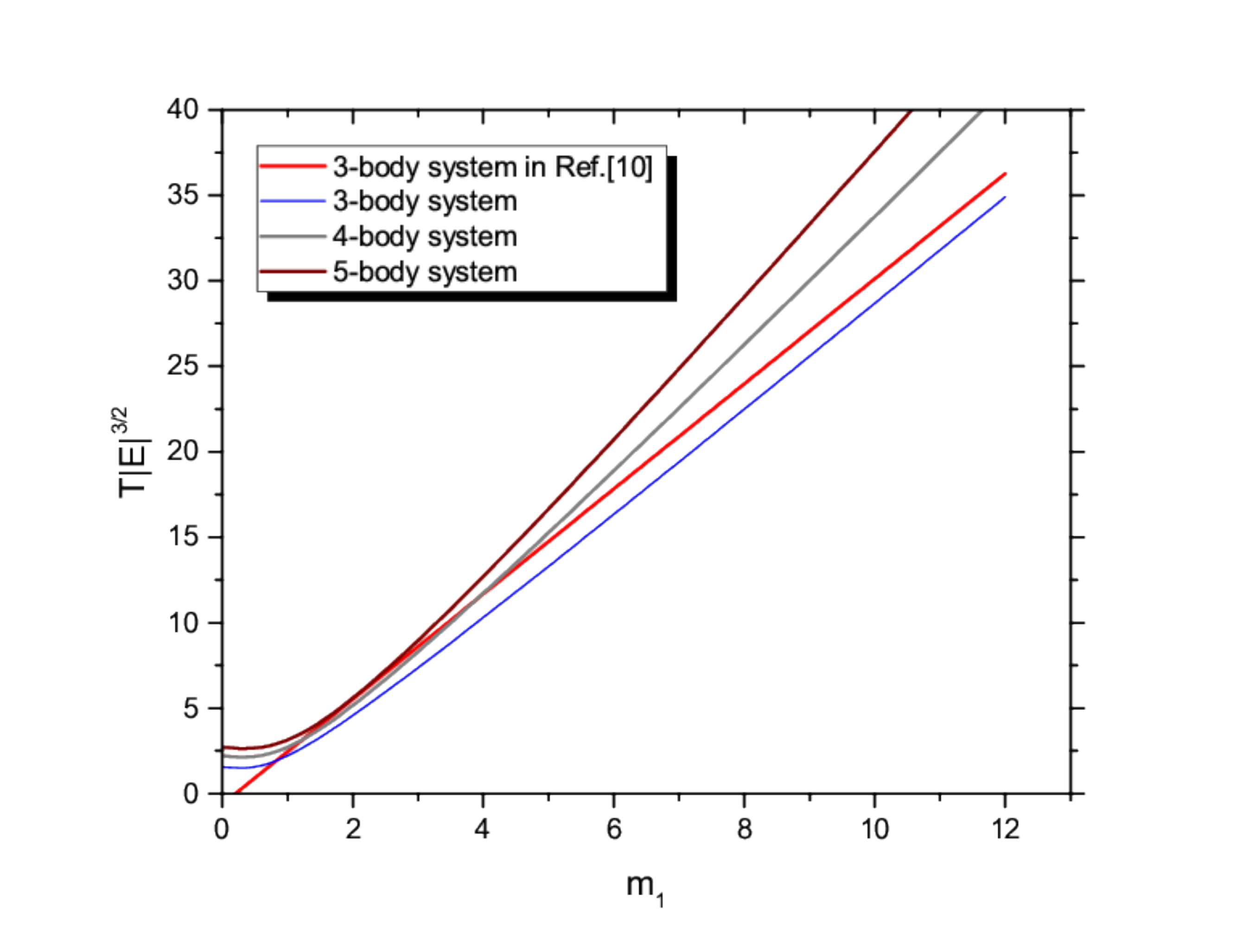}}
\caption{\label{fig5} Comparing of 3-body, 4-body and 5-body system at $m_1\in [0,12]$.}
\end{figure}

Case 5: If we keep $m_1$ varied and set all other point masses to be unit mass, namely $m_k=1,\, k\neq 1$, what's going to happen?  Kepler's third law of those four cases have been illustrated in the Fig.(\ref{fig5}), which indicate that the more point masses the system has the higher orbit period it has. In general, $T_{n+1}|E_{n+1}|^{3/2}> T_n|E_n|^{3/2}$ can be proven.

Case 6: For a system with point masses $(M, 1,1,1,..,1)$, if $M$ were massive and much heavier than other unit mass and their summation, then we have an interesting result as follows $T_n|E_n|^{3/2} \approx (\frac{n-1}{2})^{1/2}\pi G M$, clearly, it is a linear law of single massive mass, $M$.

This study considered the periodic orbital period of an n-body system from the perspective of dimensional analysis and symmetry of mass product. The universal law of an n-body system is deduced: Kepler's third law, namely the periodic law, states that periodic motion of an n-body system satisfies the $3/4$ power law of the orbital area, or the $3/2$ power law of the total energy, and or the product of $T_n|E_n|^{3/2}$ is a constant for a given point-mass system. In light of Kepler¡¯s third law, we proposed a generalized Kepler¡¯s third law for an n-body system. A numerical validation and comparison study was hence conducted. This study may open a new avenue for the investigation of the multi-body system.

I am grateful to Prof. Shijun Liao, Dr. Xiaoming Li and Michael Sun for their interest and comments in this work.


\end{document}